\documentclass[11pt]{article}
\usepackage[utf8]{inputenc}
\usepackage{caption}
\usepackage{subcaption}
\usepackage{braket,slashed,bm}
\usepackage{jheppub}
\usepackage{simplewick}
\usepackage{array,multirow}
\usepackage{amsmath} 
\usepackage[normalem]{ulem}
\usepackage{calligra}
\usepackage{floatrow}
\DeclareUnicodeCharacter{2212}{\ensuremath{-}}
\DeclareMathAlphabet{\mathpzc}{OT1}{pzc}{m}{it}
\usepackage{mathrsfs}
\usepackage{tikz}
\usepackage{rotating}
\usepackage{pgf}

\usepackage{subcaption}
\usepackage{lscape}
\usepackage{graphicx}
\usepackage{booktabs}
\newfloatcommand{capbtabbox}{table}[][\FBwidth]

\newcommand{\lsim}{\mathrel{\mathop{\kern 0pt \rlap
  {\raise.2ex\hbox{$<$}}}
  \lower.9ex\hbox{\kern-.190em $\sim$}}}
\newcommand{\gsim}{\mathrel{\mathop{\kern 0pt \rlap
  {\raise.2ex\hbox{$>$}}}
  \lower.9ex\hbox{\kern-.190em $\sim$}}}

\newcommand{\mev}{\ensuremath{\,\mathrm{MeV}}}
\newcommand{\gev}{\ensuremath{\,\mathrm{GeV}}}

\DeclareUnicodeCharacter{2212}{\ensuremath{-}}

\allowdisplaybreaks
\begin{document}

\title{
Weak Lensing Constraints on Dark Matter-Baryon Interactions with $N$-Body Simulations and Machine Learning
}

\author[a,b]{Chi Zhang,}
\author[a,1]{Lei Zu\note{Corresponding Author, email: zulei@pmo.ac.cn},}
\author[c]{Hou-Zun Chen,}
\author[a,b,2]{Yue-Lin Sming Tsai\note{Corresponding Author, email: smingtsai@pmo.ac.cn},}
\author[a,b]{Yi-Zhong Fan}

\affiliation[a]{Key Laboratory of Dark Matter and Space Astronomy,
Purple Mountain Observatory, Chinese Academy of Sciences, Nanjing 210023, People’s Republic of China}
\affiliation[b]{School of Astronomy and Space Science, University of Science and Technology of China, Hefei 230026, People’s Republic of China}
\affiliation[c]{Institute for Astronomy, the School of Physics, Zhejiang University, Hangzhou 310027, People’s Republic of China}

\abstract{
We investigate the elastic scattering cross section between dark matter and protons using the DES Year 3 weak lensing data. 
This scattering induces a dark acoustic oscillation structure in the matter power spectra. 
To address non-linear effects at low redshift, we utilize principal component analysis alongside a limited set of $N$-body simulations, 
improving the reliability of our matter power spectrum prediction. 
We further perform a robust Markov Chain Monte Carlo analysis 
to derive the upper bounds on the DM-proton elastic scattering cross-section, 
assuming different velocity dependencies. 
Our results, presented as the first Frequentist upper limits, are compared with the ones obtained by Bayesian approach. 
Compared with the upper limits derived from the Planck cosmic microwave background data, 
our findings from DES Year 3 data exhibit improvements of up to a factor of five. 
In addition, we forecast the future sensitivities of the China Space Station Telescope, 
the upcoming capabilities of this telescope could improve the current limits by approximately one order of magnitude. 
}
\date{\today}
\maketitle
\section{Introduction}

Dark Matter (DM) is one of the most fundamental mysteries in modern physics, even though its gravitational effects are well understood. 
Besides gravity, the interactions between DM and baryons are also of great interest, 
as explored by experiments such as PandaX~\cite{PandaX:2023toi} and XENONnT~\cite{XENON:2023cxc}. 
However, no signal has been detected for DM masses above about $1\gev$. 
Therefore, the focus of experiments has shifted to the sub-GeV mass range, 
for example, CDEX~\cite{CDEX:2022rxz}, SENSEI~\cite{SENSEI:2023zdf}, etc.

The interaction between DM and baryons can change the matter distribution in our universe, 
creating a dark acoustic oscillation (DAO) feature in the matter power spectra. 
DAO affects the Cosmic microwave background (CMB) anisotropies and the structure formation in the early universe, 
and constrains the DM-proton elastic scattering cross-section~\cite{Boddy:2018wzy, Slatyer:2018aqg, Boddy:2018kfv, Buen-Abad:2021mvc}. 
However, the DAO suppression, similar to that of warm DM, is more sensitive to small-scale observations. 
For velocity independent case, one obtains a stronger limit on this cross-section ($\sigma_{\chi p}<2.8 \times 10^{-28}$~cm$^2$
for DM mass around $10$~MeV) from the Milky Way satellite abundance~\cite{Nadler:2019zrb,Maamari:2020aqz,DES:2020fxi, Buen-Abad:2021mvc}, 
and the tightest limit ($\sigma_{\chi p}<1.7 \times 10^{-29}$~cm$^2$ for DM mass around $10$~MeV) from 
the Lyman-$\alpha$-forest~\cite{Hooper:2021rjc, Rogers:2021byl, Buen-Abad:2021mvc}. 
Nevertheless, the theoretical predictions of these small-scale observations are affected by the non-linear evolution of power spectra and the baryonic feedback. 
A recent research have explored such baryonic feedback in the galaxies affected by the DM-baryon interactions through hydrodynamical simulation~\cite{Acevedo:2023cab}.
However, the systematic uncertainties of these predictions remain unclear.

Weak Gravitational Lensing (WL) is a good tool for probing the late-time Large Scale Structure (LSS) of the universe. 
Through statistical analyses of shape distortions in numerous galaxies induced by foreground matter fields, 
it can directly map the LSS of the universe. 
We can mask the small-scale WL data to reduce the uncertainties associated with baryonic feedback, 
which are more pronounced in observations of the Lyman-$\alpha$-forest at the small scales.
In addition, WL data is expected to be more sensitive than the CMB anisotropies.
Many recent and upcoming surveys, including Dark Energy Survey (DES)~\cite{DES:2017myr}, 
Kilo-Degree Survey (KiDS)~\cite{vanUitert:2017ieu, Heymans:2020gsg}, 
Subaru Hyper Suprime-Cam (HSC)~\cite{HSC:2018mrq,Hamana:2019etx}, Euclid~\cite{EUCLID:2011zbd},
the Vera C. Rubin Observatory~\cite{LSST:2008ijt},
the Nancy Grace Roman Space Telescope~\cite{Green:2012mj},
the Wide Filed Survey Telescope (WFST)~\cite{Lou:2016,Lei:2023adp}, 
the Mephisto Telescope~\cite{Lei:2021vtu,2019gage.confE..14L} and China Space Station Telescope (CSST)~\cite{Zhan:2011,Zhan:2021,Gong:2019yxt}, 
greatly improve our understanding of the matter distribution in the late universe. 
They, in turn, have the potential to reveal the fundamental physics of the interaction between DM and baryonic matter.
In this work, we use the data from DES three-year (DES~Y3) `\texttt{3$\times$2pt}' WL observations along with the CMB and baryonic acoustic oscillation (BAO) observations data.
The `\texttt{3$\times$2pt}' WL observations include three set of correlation functions: (i) cosmic shear, the shape-shape correlations of the source galaxies; (ii) galaxy clustering, the position-position correlations of the lens galaxies; (iii) galaxy-galaxy lensing, the cross correlations between source shape and lens positions.
Furthermore, we generate the mock data for CSST and present a forecast of the power of CSST. 

Because the photometric galaxy surveys can cover the red-shift with the range $0<z<5$,
the non-linear effects on the matter power spectrum are essential for the theoretical prediction of WL signal.
In this study, we conducted a series of DM-only $N$-body simulations to accurately account for the non-linear effects on the matter power spectrum. 
The matter power spectrum can be modified by the elastic scattering of DM particle $\chi$ with proton $p$. 
The scattering cross-section, denoted as $\sigma_{\rm \chi p}\equiv \sigma_n v_{\rm rel.}^n$, is parameterized by a power-law index $n$ 
and the relative velocity between DM and protons $v_{\rm rel.}$ in units of the speed of light. 
These scattering interactions induce perturbations in the CMB power spectrum, leading to suppression and oscillations in the matter power spectrum, 
as well as higher-order statistical quantities, e.g. bispectrum.
Taking into account the nonlinear effects from gravitation and scattering, especially at the suppression scale $k \gsim 0.1\,h/\rm{Mpc}$, is crucial. 
However, it is a great challenge to incorporate DM-proton scattering in cosmological hydrodynamic simulations, particularly for negative values of $n$ 
that may govern the nonlinear effects in the late-time Universe. 
In the case of $n \geqslant 0$, because $v_{\rm rel}$ in the late-time Universe is also suppressed, 
the DM-proton scattering cross-section can be ignored in cosmological hydrodynamic simulations. 
Hence, for simplicity, we only discuss the $n \geqslant 0$ scenario to address the nonlinear effects at low redshift. 
To fast compute our matter power spectrum prediction at low redshift, 
we employ principal component analysis (PCA) together with a limited set of $N$-body simulations.

The structure of this paper is as follows: We first summarize the linear evolution in DM-proton scattering in Sec.~\ref{theroy}.
In Sec.~\ref{methodology}, we present our approach to a fast computation of non-linear correction including DM-proton scattering. 
In Sec.~\ref{MCMC}, we conduct a high-dimensional Markov Chain Monte Carlo (MCMC) scan to explore the parameter space. 
Our results are presented and discussed in Sec.~\ref{results}. 
Finally, we summarize and conclude in Sec.~\ref{conclusion}.

\section{Linear evolution} 
\label{theroy}

\begin{figure}[h!]
    \centering
    \includegraphics[scale=0.45]{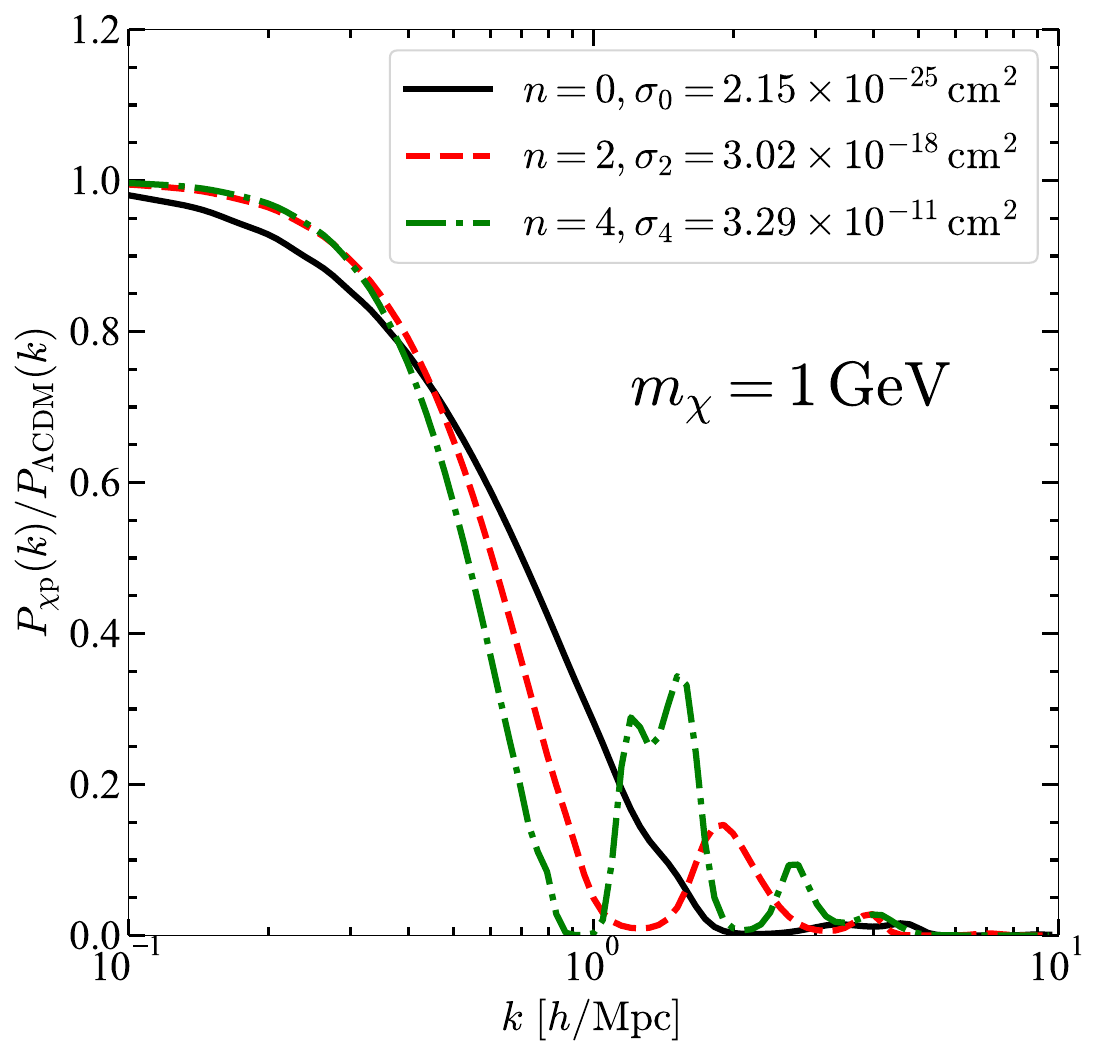}
    \caption{The ratio of the linear matter power spectrum in DM-proton scattering scenario $P_{\rm \chi p}(k)$ to 
    that in $\Lambda$CDM scenario $P_{\Lambda \rm{CDM}}(k)$. 
    The black solid line, red dashed line, and green dashed-dotted line correspond to $n=0$, $n=2$, and $n=4$ at $m_\chi=1\,\rm{GeV}$, respectively. 
    The cross sections $\sigma_{n=0,2,4}$ are the $95\%$ upper limits obtained by P18 + BAO + DES \texttt{3$\times$2pt} likelihood in Sec.~\ref{results}.}
    \label{fig:compare_n}
\end{figure}

Considering the elastic scattering interactions between DM ($\chi$) and baryon ($b$), 
a collision term emerges within the standard Boltzmann equation~\cite{Chen:2002yh,Dvorkin:2013cea,Nguyen:2021cnb}, 
\begin{equation}
\begin{aligned}
& \dot{\delta}_b=-\theta_b-\frac{\dot{h}}{2}, \\
& \dot{\theta}_b=-\frac{\dot{a}}{a} \theta_b+c_b^2 k^2 \delta_b+R_\gamma\left(\theta_\gamma-\theta_b\right)+\frac{\rho_\chi}{\rho_b} R_\chi\left(\theta_\chi-\theta_b\right), \\
& \dot{\delta}_\chi=-\theta_\chi-\frac{\dot{h}}{2}, \\
& \dot{\theta}_\chi=-\frac{\dot{a}}{a} \theta_\chi+c_\chi^2 k^2 \delta_\chi+R_\chi\left(\theta_b-\theta_\chi\right).
\end{aligned}
\end{equation}
The density fluctuations, the velocity divergences in Fourier space, 
the energy density, and 
the speeds of sound in each fluid are denoted as $\delta_i$, $\theta_i$, $\rho_i$, and $c_{i}$, respectively, 
where subscript $i$ is $b$ for baryon but $\chi$ for DM. 
The over-dot represents a derivative for conformal time.
The quantity $k$ is the wave number, $a$ is the scale factor, and $h$ is the trace of the scalar metric perturbation.
The coefficients $R_{\chi}$ and $R_{\gamma}$ represent the momentum-transfer rate of DM-proton scattering and the standard Compton scattering.
The temperatures of baryon and DM are
\begin{equation}
\begin{aligned}
& \dot{T}_b+2 \frac{\dot{a}}{a} T_b=2 \frac{\mu_b}{m_e}
R_\gamma\left(T_\gamma-T_b\right)+2 \frac{\mu_b}{m_\chi} \frac{\rho_\chi}{\rho_b} R_\chi^{\prime}\left(T_\chi-T_b\right) \\
& \dot{T}_\chi+2 \frac{\dot{a}}{a} T_\chi=2 R_\chi^{\prime}\left(T_b-T_\chi\right),
\end{aligned}
\end{equation}
where the temperature of baryon, photon, and DM are $T_b$, $T_\gamma$, and $T_\chi$.
The quantity $m_e$ is the mass of the electron, $m_\chi$ is the mass of DM particle, $\mu_b$ is the mean molecular weight of the baryons. 
The heat transfer rate coefficient $R_\chi^\prime = R_\chi m_\chi/(m_\chi+m_p)$.
The DM-proton scattering cross section $\sigma_{\chi p}$ can be simply parameterized 
as a power law of the DM-proton relative velocity $v_{\rm rel.}$, 
\begin{equation}
    \sigma_{\chi p} = \sigma_n v_{\rm rel.}^n,  
    \label{eq:Xsec}
\end{equation}
where $\sigma_n$ is a dimensional constant factor and the index $n$ can be either a positive or a negative integer. 
In terms of velocity dependence, the momentum transfer rate coefficient $R_\chi$ can be written as 
\begin{equation}
R_\chi=a \rho_b \frac{Y_\mathrm{H} \mathcal{N}_n \sigma_n}{m_\chi+m_p}\left(\frac{T_\chi}{m_\chi}+\frac{T_b}{m_p}\right)^{(n+1) / 2},
\label{eq:Rchi}
\end{equation}
where the mass fraction of hydrogen is $Y_{\mathrm{H}}$ and the mass of a proton is $m_p$. 
The numerical factor $\mathcal{N}_n \equiv 2^{(5+n) / 2} \Gamma(3+n / 2) /(3 \sqrt{\pi})$. More details are shown in~\cite{Dvorkin:2013cea}. 
Note that we only consider the positive cases in this work, namely $n=0,2,4$~\cite{Chen:2002yh,Sigurdson:2004zp,Melchiorri_2007}, 
but will return to $n<0$ cases in the future.
Notably, for the case of $n \geq 0$, the bulk relative velocities are much smaller in comparison to the thermal velocities, as mentioned in \cite{Dvorkin:2013cea,Ali-Haimoud:2023pbi}.
We neglect the impact of bulk relative velocities here and thus the calculation in \cite{Ali-Haimoud:2023pbi} does not affect our analysis.

To model the effects of DM-proton interactions on both the linear matter power spectra and the CMB power spectra, 
we use a modified version of the cosmological Boltzmann solver \texttt{CLASS}~\cite{Blas:2011rf,Nguyen:2021cnb}.
In Fig.~\ref{fig:compare_n}, we present the ratios of the linear matter power spectra 
in the DM-proton scattering scenario $P_{\rm \chi p}$ to that in the $\Lambda$CDM $P_{\Lambda {\rm CDM}}$ for a DM mass $m_\chi=1\,\rm{GeV}$. 
We present three velocity dependencies: $n=0$ (the black solid line), $n=2$ (the red dashed line), and $n=4$ (the green dashed-dotted line). 
Despite varying cross-sections, the power spectra are all suppressed at $k\sim 0.3~h/$Mpc. 
These similar suppression are because we take the cross-sections from their $95\%$ upper limits obtained by P18 + BAO + DES \texttt{3$\times$2pt} likelihood,  
whose detailed description will be given in Sec.~\ref{results}. 
The velocity dependencies show distinct oscillations around $k\sim 1~h/$Mpc. 
However, since the predicted WL signals are obtained by integrating over $k$ modes, 
the tested likelihood is less sensitive to the phase of oscillations comparing with the overall suppression in matter power spectrum.

\section{Non-linear matter power spectra}
\label{methodology}

\begin{figure}[ht!]
    \centering
    \includegraphics[scale=0.40]{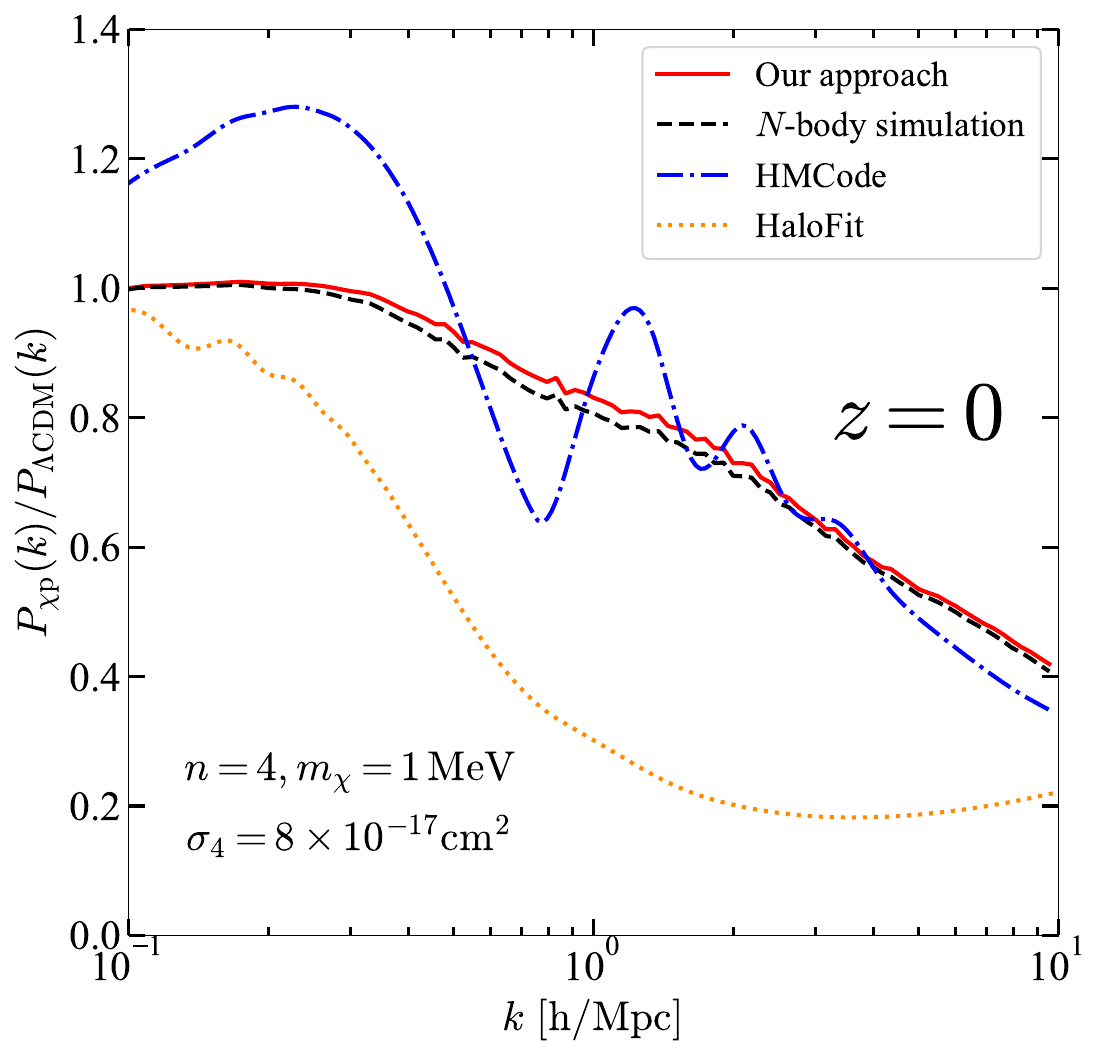}
    \includegraphics[scale=0.40]{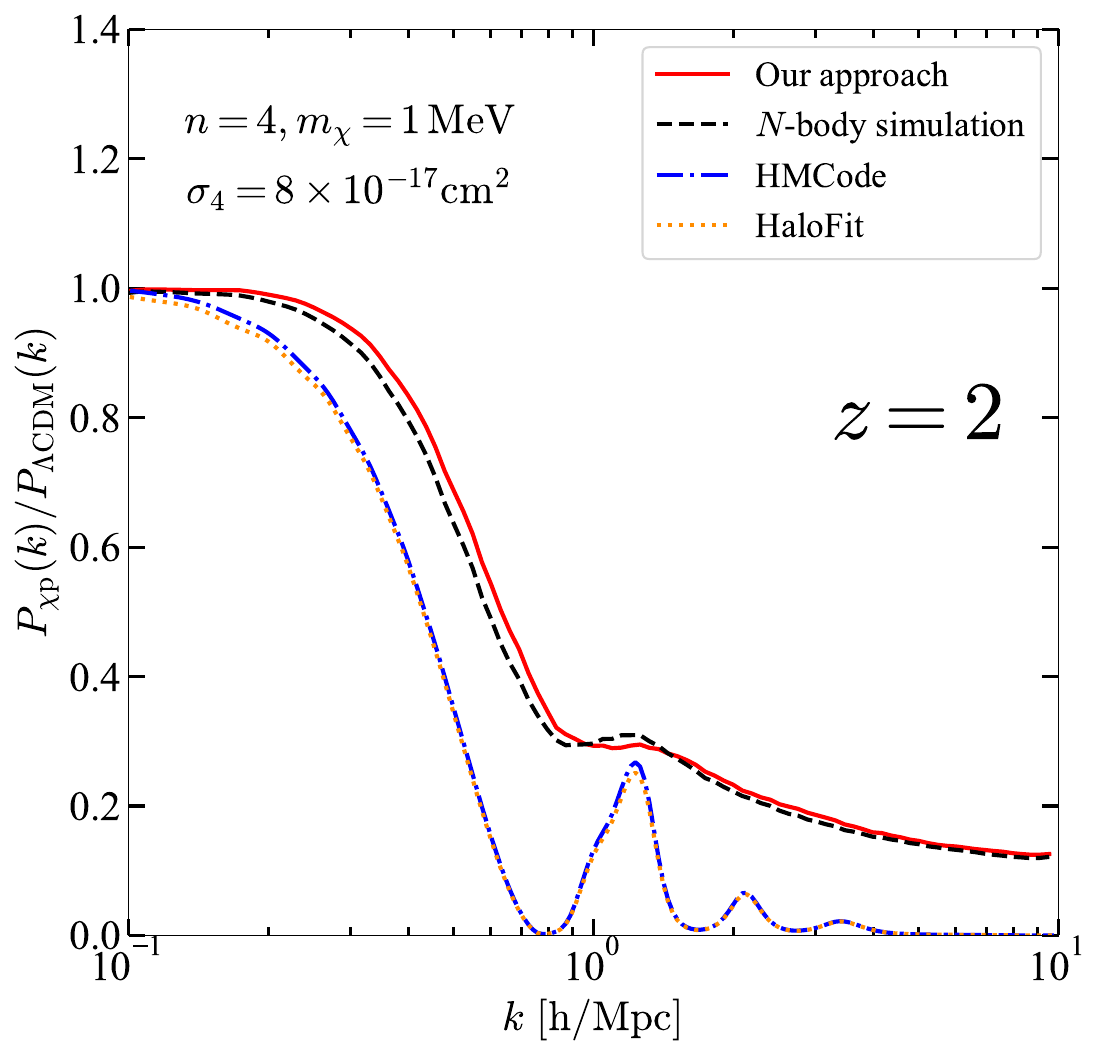}
    \caption{
    The left (right) panel corresponding to the ratio of the non-linear matter power spectrum at $z=0$ ($z=2$) 
    with example parameters $n=4$, $m_\chi=1\rm{MeV}$, $\sigma_4=8 \times 10^{-17} \rm{cm}^2$. 
    The red solid lines denote the outputs of our approach, the black dashed lines are the $N$-body simulation results, the blue dashed-dotted lines are the correction from \texttt{HMCode} and the orange dotted lines are the correction from \texttt{HaloFit}.
    }
    \label{fig:simu_dmp_pks}
\end{figure}
The linear perturbation theory relies on the assumption that the density fluctuation $\delta(k)$ remains significantly smaller than unity. 
However, during cosmic structure formation, the growth of matter density fluctuations can result in $\delta(k) \gg 1$, rendering it no longer a perturbation.
As a consequence, the linear perturbation theory becomes inadequate for describing the intricate structural evolution 
in regions with wave numbers $k \gtrsim 0.1~h/\rm{Mpc}$, commonly referred to as the non-linear region. 
In addition, DM-proton interactions with a cross-section characterized by $n>0$ cease to contribute to non-linear evolution, 
as their interaction rate falls below the Hubble expansion rate. 

The easiest but computationally expensive way to explore the non-linear region is to conduct $N$-body simulations. 
To efficiently include accurate non-linear corrections,  
several research groups have developed varied analytical methodologies to rapidly compute non-linear corrections.
For example, widely used tools \texttt{HMCode}~\cite{Mead_2016} and \texttt{Halofit}~\cite{Takahashi_2012} rely on classical CDM simulation outcomes, 
as well as \texttt{WarmAndFuzzy}~\cite{Marsh:2016vgj} rely on warm dark matter (WDM) and fuzzy dark matter (FDM) simulations.
In this work, we develop a new approach to speed up our non-linear matter power spectrum computations, which 
allows us to reuse $\mathcal{O}(100)$ $N$-body simulations for a global scan.    

First, we define the matter power spectrum ratio $\mathcal{R}$ between $P_{\Lambda {\rm CDM}}$ and $P_{\rm \chi p}$, similar to the method applied in~\cite{Cataneo:2018cic},
\begin{equation}
    \mathcal{R}^{i}(k,z) \equiv 
    \frac{P^{i}_{\rm \chi p}(k,z)}{P^{i}_{\Lambda \rm{CDM}}(k,z)},
\label{eq:R}
\end{equation}
where $i$ is either ``nl" or ``lin" for the cases with and without non-linear effect. 
To simplify the expression, here we omit the standard arguments ($\Lambda$CDM cosmological parameters, $m_\chi$, 
and $\sigma_n$) in Eq.~\ref{eq:R}.   
We compute $\mathcal{R}^{\rm lin}$ 
by using the Boltzmann code \texttt{CLASS}~\cite{Blas:2011rf}, a modified version including DM-proton scattering~\cite{Nguyen:2021cnb}, while  
$\mathcal{R}^{\rm nl}$ is computed by using $N$-body simulation code \texttt{GIZMO}~\cite{Hopkins:2014qka}. 
In the Appendix~\ref{nbody_set}, we provide a detailed description of our $N$-body simulations.
Note that $\mathcal{R}$ can be strongly dependent on a few parameters. 
Hence, we then use the widely adopted machine learning skill, principal component analysis (PCA), 
to investigate how many free parameters there are.
For $n=0,2,4$ with DM mass between 1 MeV to 1 TeV, 
we scrutinize the parameter space suggested by Ref.~\cite{Li:2022mdj} 
to verify that the PCA machine can precisely interpolate 
the linear ratios $\mathcal{R}^{\rm lin}(k,z)$ 
by two parameters~\footnote{A similar method has been utilized 
to investigate two-body decaying DM in ~\cite{Bucko:2023eix}.}. 
Such a collected set of $\mathcal{R}^{\rm lin}$ generated by parameter scan can be then used 
for training and validation.  
After training our PCA machine, we can simply use it to compress any $\mathcal{R}^{\rm lin}$ to only two PCA components.

Our next goal is to build reusable $\mathcal{R}^{\rm nl}$ tables as function of two PCA components. 
We uniformly select 205 points within this two-dimensional map from previously collected training and validating data. 
We then perform 205 $N$-body simulations accordingly to obtain corresponding $\mathcal{R}^{\rm nl}$. 
Consequently, for any linear power spectrum falling within our specified range of interest, a straightforward mapping from linear to non-linear matter power spectrum becomes achievable. 
Namely, we can interpolate the $\mathcal{R}^{\rm nl}$ tables with their corresponding two PCA components. 
Hence, the final non-linear matter power spectrum for DM-proton scattering can be obtained via
\begin{equation}
    \label{halofit_times_ratio}
    P^{\rm nl}_{\rm \chi p}(k,z) = \mathcal{R}^{\rm nl}(k,z) \times P_{\rm \Lambda CDM}^{\texttt{Halofit}}(k,z),
\end{equation}
where $P_{\rm \Lambda CDM}^{\texttt{Halofit}}(k,z)$ is the non-linear matter power spectrum of $\Lambda$CDM cosmology computed by \texttt{Halofit}.
We emphasize that the $N$-body simulations are only dependent on initial conditions making our approach applicable to various DAO models. 
This is the greatest advantage of our method, even though this paper only focuses on the DM-proton scattering scenario.

In Fig.~\ref{fig:simu_dmp_pks}, we present a comparison between 
our approach (red solid lines),
$N$-body simulations (black dashed lines), 
\texttt{HMCode} (blue dashed-dotted lines), 
and \texttt{HaloFit} (orange dotted lines) 
for a benchmark point ($n=4$, $m_\chi=1~\rm{MeV}$, $\sigma_4=8 \times 10^{-17} \rm{cm}^2$). 

For both $z=0$ (left panel) and $z=2$ (right panel), 
we can see that our approach predicts the $N$-body simulation well.  
On the other hand, the \texttt{HMCode} and \texttt{HaloFit} cannot catch the features of the non-linear effects, 
because they exclusively rely on $\Lambda$CDM simulations.
Quantitatively, our approach generates an acceptable error around $0.5\%$.  
For more information, the error estimation is given in Appendix~\ref{error}.

\section{Markov Chain Monte Carlo analysis}
\label{MCMC}

Based on PCA machine and the $\mathcal{R}^{\rm nl}$ tables developed in Sec.~\ref{methodology}, 
we can include non-linear corrections on the fly when performing a MCMC scan. 
The scan parameters are the six fundamental $\Lambda$CDM cosmological parameters, 
and $\sigma_{n=0,2,4}$ as given in Eq.~\ref{eq:Xsec}.  
The six $\Lambda$CDM parameters consist of
\begin{equation}
  \{\Omega_{\rm b} h^2,~\Omega_{\rm cdm} h^2,~\theta_{\rm s},~\log(10^{10} A_{\rm s}),~n_{\rm s},~\tau_{\rm reio}\}, 
  \label{eq:cosmo_param}
\end{equation}
corresponding to the baryon density, the cold DM density, the CMB peak position, the scalar amplitude, 
the scalar spectral index, and the optical depth to reionization, respectively. 
The scan range required for Eq.~\ref{eq:cosmo_param} can be found in previous work~\cite{Zu:2023rmc}. 
We assume all the DM particles can scatter with proton.
Except for $\sigma_n$, we select seven benchmark masses $m_\chi \in \{10^{-3}, 10^{-2}, 10^{-1}, 10^0, 10^1, 10^2, 10^3\}$~GeV, and 
the three power indexes $n = 0, 2, 4$ for different scans. 
Namely, we have $7\times 3$ scans in total for those benchmark masses and power indexes.  
We employ the cosmological MCMC package~\texttt{MontePython}~\cite{Audren:2012wb,Brinckmann:2018cvx} to undertake the task of the global fitting. 

We include the following cosmological data into the likelihood: 
(i) The \texttt{3$\times$2pt} likelihood based on the DES~Y3 observations~\cite{DES:2021wwk,Zu:2023rmc}, 
(ii) The CMB likelihoods are calculated based on Planck 2018 Legacy (P18)~\cite{Planck:2019nip}, 
including high-$\ell$ power spectra (\texttt{TT}, \texttt{TE}, and \texttt{EE}), 
low-$\ell$ power spectrum (\texttt{TT} and \texttt{EE}), and Planck lensing power spectrum (\texttt{lensing}), 
(iii) The BAO likelihood contains the BOSS DR12 dataset measurements at $z=0.106$, $z=0.15$ and $z=0.2-0.75$~\cite{Beutler_2011,Ross_2015,Alam_2017}.
The details of DES~Y3 \texttt{3$\times$2pt} data and modeling are given in the Appendix~\ref{WL}.

\section{Results and discussion}
\label{results}

In this study, we determine the $95\%$ upper limit of $\sigma_n$ using two methods: Bayesian Marginalized Posterior (MP) and Frequentist Profile Likelihood (PL). 
Both methods set upper limits by integrating $\sigma_n$ until $95\%$ of the total probability is reached. 
The method MP integrates the probability density over nuisance parameters by marginalizing posterior densities, commonly applied in cosmology, particularly in the $\Lambda$CDM context. 
On the other hand, the PL method is preferred for null signal searches due to large volume effects and prior dependencies in unconstrained likelihoods, 
and it is used in DM direct detection~\cite{PandaX:2023toi,XENON:2023cxc,CDEX:2022rxz} and indirect detection~\cite{Fermi-LAT:2016uux,PhysRevD.104.083026,Hess:2021cdp}. 

In our study on DM-proton scattering cross-section, we primarily present upper limits using the PL method, with results from the MP method included for comparative analysis~\footnote{Recently, Ref.~\cite{Fowlie:2024dgj} has proposed that the Bayes factor surface may be an alternative good tool for dealing with null signal search.}.

\begin{figure*}[ht!]
    \centering
    \includegraphics[scale=0.45]{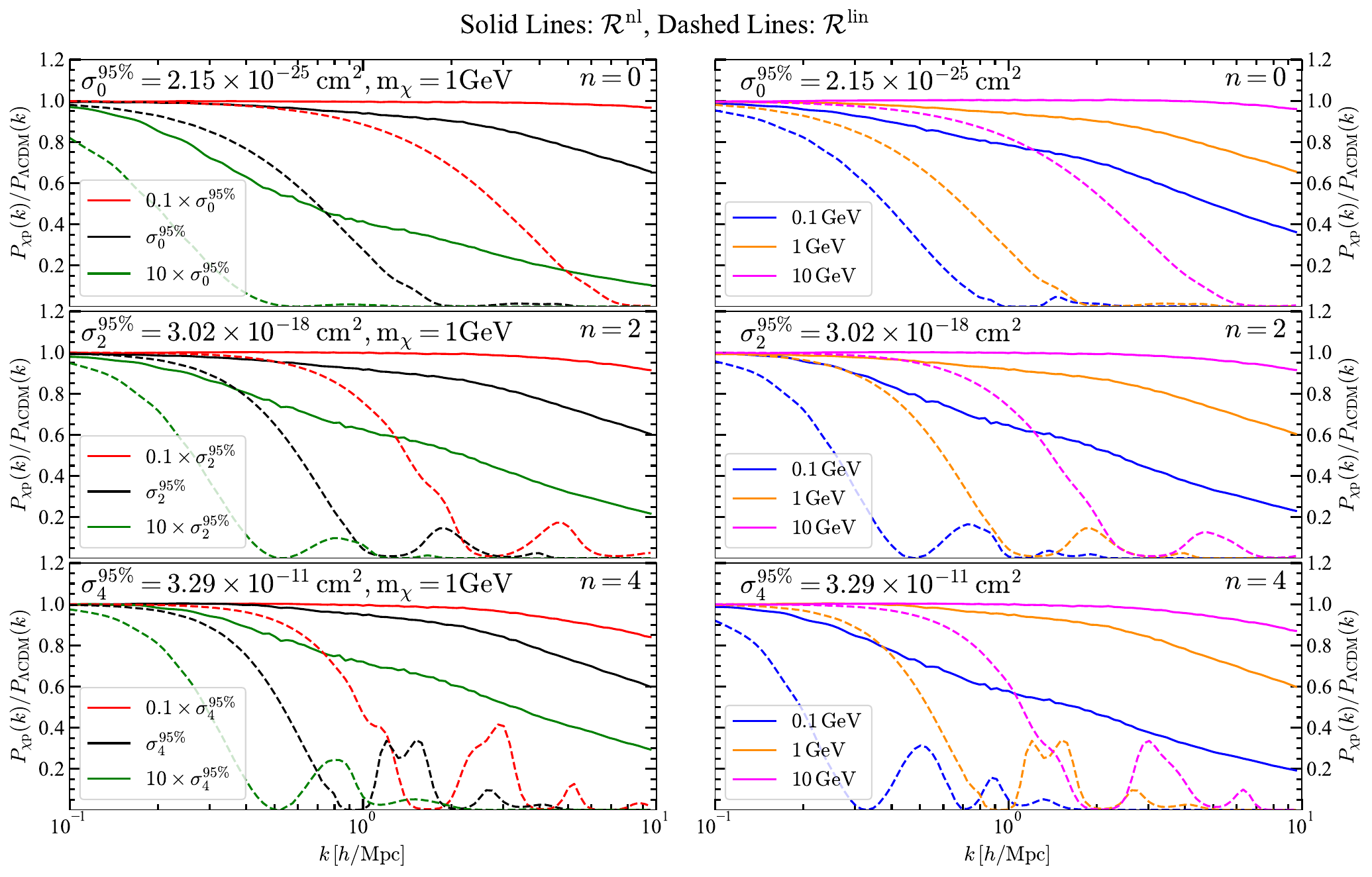}
    \caption{The ratio power spectra caused by the DM-proton scatterings. 
    The dashed lines are the linear results, and the solid lines are the non-linear results.
    Left Column: the ratio power spectra for DM mass $m_\chi=1\gev$ with the velocity scenarios $n=0$ (left upper panel), 
    $n=2$ (left middle panel), and $n=4$ (left lower panel). 
The reference cross-sections are $\sigma^{\mathrm{95\%}}_0 = 2.15 \times 10^{-25}\, \mathrm{cm}^2$, $\sigma^{\mathrm{95\%}}_2 = 3.02 \times 10^{-18}\, \mathrm{cm}^2$, and $\sigma^{\mathrm{95\%}}_4 = 3.29 \times 10^{-11}\, \mathrm{cm}^2$. These values represent the $95\%$ upper limits based on the likelihood P18 + BAO + DES \texttt{3$\times$2pt}, as discussed in Sec.~\ref{results}.
    Right column: the ratio spectra for DM mass $m_\chi=100\mev$ (blue lines), $m_\chi=1\gev$ (orange lines) and $m_\chi=10\gev$ (magenta lines) for $n=0,2,4$, with the same cross-section given in the right panels. 
    }
    \label{fig:6fig}
\end{figure*}

In Fig.~\ref{fig:6fig}, we present the ratio spectra at $z=0$ as a function of $k$, by varying $\sigma_{\chi p}$ in the left panels and $m_\chi$ in the right panels.

In the left panels, we take a constant DM mass at $m_\chi=1~\mathrm{GeV}$ and investigate three different velocity scenarios: $n=0$ (upper left panel), $n=2$ (middle left panel), and $n=4$ (lower left panel). 
Three colored lines correspond to the cross-sections $\sigma_{n=0,2,4}$ set at $0.1 \times \sigma_n^{\rm 95\%}$ (the red lines), $\sigma_n^{\rm 95\%}$ (the black lines), and $10 \times \sigma_n^{\rm 95\%}$ (the green lines).  
When comparing the non-linear and linear ratio spectra (represented by solid and dashed lines, respectively), 
the non-linear effects tend to washout the features of DAO and cause a shift in suppression towards smaller scales. 
The values of $\sigma_{n=0,2,4}^{\rm 95\%}$ displayed in the left corner of each figure are derived from the $95\%$ upper limits of the likelihood (P18 + BAO + DES \texttt{3$\times$2pt}). 
These values can be considered as characteristic cross-sections. 
For the sake of convenience, we subsequently define $k_{0.8}$ as the wavenumber where the ratio spectrum is measured at $0.8$. 
Observing all non-linear ratio spectra with $\sigma_{n=0,2,4}^{\rm 95\%}$, we find that $k_{0.8}$ is approximately $5~h/\rm{Mpc}$. 
This value also indicates the sensitive scale of the DES \texttt{3$\times$2pt} data.

An important trend emerges as reducing the cross-section value leads to the suppression of the matter power spectrum, shifting it toward smaller scales. 
This shift results from an earlier decoupling between DM and baryons due to reduced interaction between them. Therefore, for future experiments aiming to enhance detection sensitivity, efforts should be directed towards smaller-scale regions, such as $k>1~h/\rm{Mpc}$. However, theoretical challenges, including baryonic feedback and galaxy bias~\cite{Desjacques:2016bnm}, introduce substantial and complex systematic uncertainties in WL surveys at small scales.

Here we discuss the assumption of a linear galaxy bias model at large scales and the scale-cuts adopted at small scales in both the DES analysis~\cite{DES:2021bvc} and also in our study.
The galaxy bias denotes the statistical relation between the distribution of galaxies and matter. 
According to Ref.~\cite{Desjacques:2016bnm}, this bias can be expanded into a series, where the dominant term is the linear component, which is scale-independent. 
However, it is known that this linear bias assumption may not hold at small scales~\cite{DES:2020rlj}, 
yet it remains applicable at large scales as demonstrated by DES within the framework of $\Lambda$CDM~\cite{DES:2021bvc}. 
As an additional analysis, we present our results based solely on the likelihood incorporating DES cosmic shear data in Appendix~\ref{CS} for comparison.

Going beyond $\Lambda$CDM, even with new interactions between DM particles and protons, characterized by a small rate $\sigma_0/m_\chi\lesssim \mathcal{O}(10^{-25})~{\rm cm}^2/\gev$, there exists an early kinetic decoupling at an epoch around $z\sim 10^4$, much earlier than galaxy formation. 
Hence, the galaxy bias in this scenario does not deviate from that of $\Lambda$CDM, allowing for the secure use of a scale-independent bias at large scales.
For analogous reasons, we also do not incorporate this interaction into our $N$-body simulations, instead of using proper initial conditions derived from modified linear evolution. 
Thus, building on previous studies~\cite{DES:2021bvc} and implementing masks on corresponding small angular scales in WL data, as described in further detail in Appendix~\ref{WL}, aims to mitigate these uncertainties. 
This strategy also enables us to extend our analysis to future WL data.

In the right panels, we vary the parameter $m_\chi$ while keeping $\sigma_n$ fixed at $\sigma_n^{95\%}$. 
The ratio spectra for $m_\chi=100\mev$ (blue lines), $m_\chi=1\gev$ (orange lines), and $m_\chi=10\gev$ (magenta lines) are compared for scenarios of $n=0,2,4$. 
Because the lighter DM results in a higher DM number density, the DM-proton interaction rate is also larger. 
This leads to the suppression of the matter power spectra for lighter DM shifting toward larger scales. 
When comparing with the left panels, we see that variations in the cross-section have a more significant impact on the location of $k_{0.8}$ than changes in the DM mass.

\begin{figure}[ht!]
    \centering
    \includegraphics[scale=0.45]{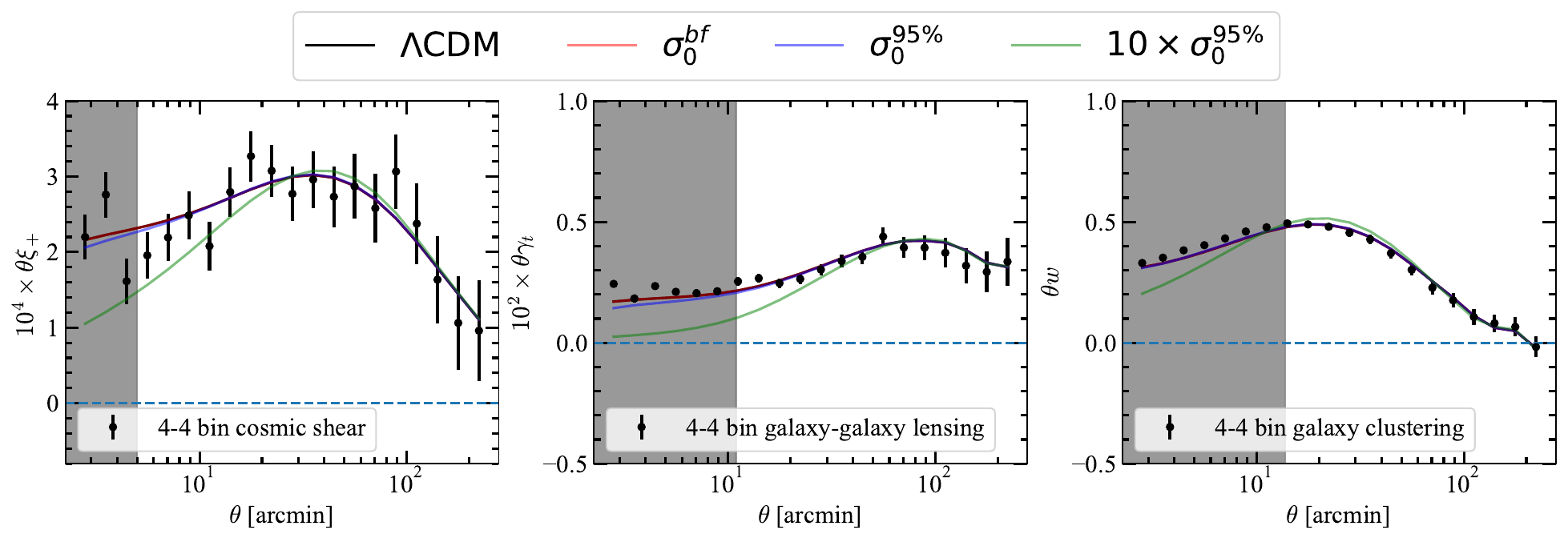}
    \caption{
    The WL 3$\times$2pt data and model predictions for CDM and DM-proton scattering with $n=0, m_\chi=1~\mev$ at the 4th-4th bin. 
    Only the 4th-4th bin is plotted for cosmic shear (source-source; left panel), galaxy-galaxy lensing (source-lens; middle panel), 
    and galaxy clustering (lens-lens; right panel). 
    The WL 3$\times$2pt data points are presented as the black error bars, while 
    the gray region is masked for the unknown systematic uncertainties.
    }
    \label{fig:3x2pt_fitting}
\end{figure}

The impact of DM-proton scattering on the WL spectrum is illustrated in Fig.~\ref{fig:3x2pt_fitting}, which presents $3\times 2$pt data and four theoretical prediction curves for a benchmark scenario ($n=0, m_\chi=1\mev$) in the 4th-4th bin. 
The figure includes cosmic shear (left panel), galaxy-galaxy lensing (middle panel), and galaxy clustering (right panel). 
The gray region is masked for the unknown systematic uncertainties.
Complete $3\times 2$pt measurements, accounting for all auto and cross correlations, are detailed in Appendix~\ref{WL}. 
In each panel, we analyze the WL signal for four typical cross-section values: 
$\sigma_0 = 0$ (representing $\Lambda$CDM, shown as the black line), 
$\sigma_0 = \sigma_0^{\rm bf}$ (red line), 
$\sigma_0 = \sigma_0^{95\%}$ (blue line), and 
$\sigma_0 = 10 \times \sigma_0^{95\%}$ (green line). 
Note that DM-proton scattering suppresses the matter power spectrum at small scales, also leading to a suppression in the predicted WL signals.

\begin{figure}[ht!]
    \centering
    \includegraphics[scale=0.28]{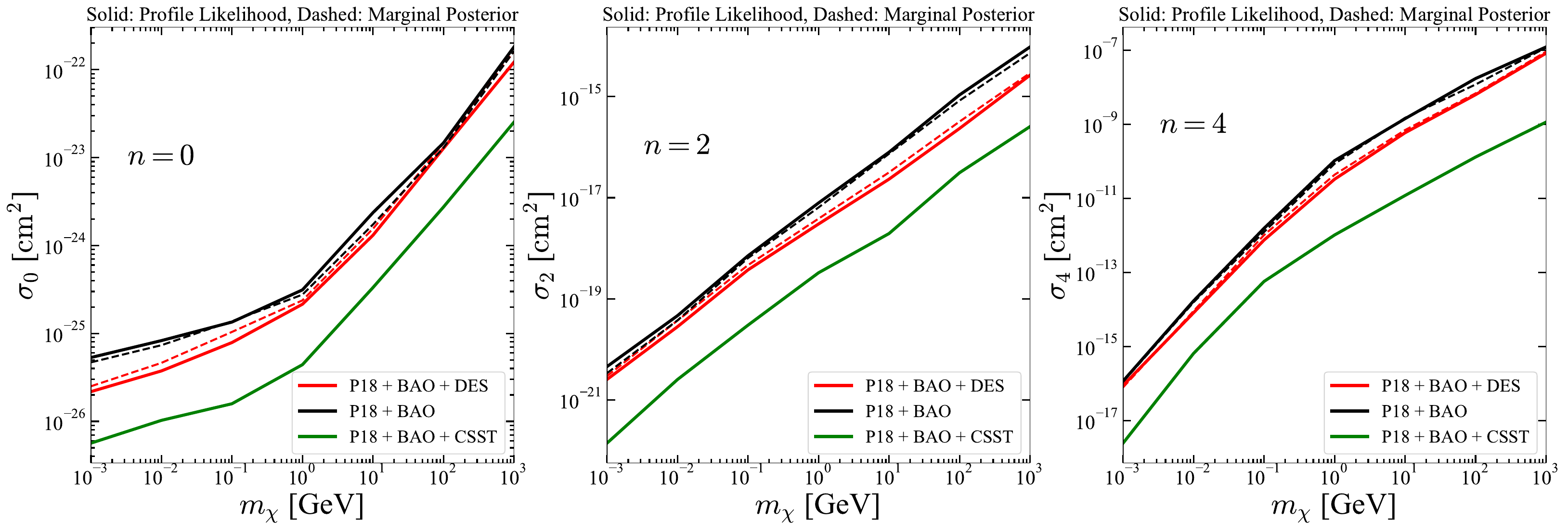}
    \caption{
    The $95\%$ upper limits of cross sections $\sigma_n$ derived by the P18 + BAO + DES likelihoods (red lines) and by the P18 + BAO likelihoods (black lines).
    The solid lines are obtained by Profile Likelihood method and the dashed lines are obtained by Marginal Posterior method.
    The green lines correspond to the forecast bounds of CSST obtained by Profile Likelihood method.}
    \label{fig:upperlimits}
\end{figure}

In Fig.~\ref{fig:upperlimits}, we present the 95\% upper limits for the DM-proton velocity-dependent elastic scattering cross-section $\sigma_n$. 
These limits are derived from likelihood functions detailed in Sec.~\ref{MCMC}, utilizing both the PL method (solid lines) and MP method (dashed lines). 
Through all velocity-dependent scenarios, the PL method establishes the most stringent upper limits, represented by red solid lines, based on combined likelihoods from \textbf{P18 + BAO + DES 3$\times$2pt}. 
Black lines denote upper limits derived solely from \textbf{P18 + BAO} likelihoods. 
Finally, we conducted a forecast to evaluate the potential constraints of upcoming CSST on DM-proton scattering, depicted by green lines.

The numerical values associated with these likelihoods used in Fig.~\ref{fig:upperlimits} are summarized in Table~\ref{tab:planck_table}, 
where \textbf{P18 + BAO + DES 3$\times$2pt} and \textbf{P18 + BAO} are detailed in the upper and lower columns respectively. 
It should be noted that for small cross-sections, the likelihood distributions are flat, 
potentially leading to arbitrary best-fit $\sigma_{0,2,4}$. 
Nonetheless, the upper limits of these cross-sections are uniquely determined.
Furthermore, Table~\ref{tab:bf_value} lists the best-fit and averaged cosmological parameters and their associated statistical strengths based on the DM parameters $\{n=0, m_\chi=1\mev\}$ for reference.

\begin{table}
\centering
\begin{tabular}{|c|c|c|c|c|c|c|}
\hline 
\multirow{2}{*}{\begin{tabular}[c]{@{}c@{}}DM mass\\ $\rm{m}_\chi$\end{tabular}} & \multicolumn{2}{c|}{$\sigma_0\left[\mathrm{~cm}^2\right]$} & \multicolumn{2}{c|}{$\sigma_2\left[\mathrm{~cm}^2\right]$} & \multicolumn{2}{c|}{$\sigma_4\left[\mathrm{~cm}^2\right]$} \\
\cline{2-7} & best-fit & upper limit & best-fit & upper limit & best-fit & upper limit \\
\hline \multicolumn{7}{|c|}{$\mathrm{P} 18+\mathrm{BAO}+$ DES $3 \times 2 \mathrm{pt}$} \\
\hline 1 MeV & $9.19 \times 10^{-28}$ & $2.18 \times 10^{-26}$ & $3.26 \times 10^{-22}$ & $2.54 \times 10^{-21}$ & $4.47 \times 10^{-20}$ & $8.92 \times 10^{-17}$ \\
\hline 10 MeV & $1.08 \times 10^{-27}$ & $3.75 \times 10^{-26}$ & $1.09 \times 10^{-20}$ & $2.80 \times 10^{-20}$ & $4.39 \times 10^{-17}$ & $8.17 \times 10^{-15}$ \\
\hline 100 MeV & $2.27 \times 10^{-27}$ & $7.88 \times 10^{-26}$ & $1.41 \times 10^{-22}$ & $3.74 \times 10^{-19}$ & $5.38 \times 10^{-13}$ & $7.51 \times 10^{-13}$ \\
\hline 1 GeV & $6.75 \times 10^{-27}$ & $2.15 \times 10^{-25}$ & $2.00 \times 10^{-21}$ & $3.02 \times 10^{-18}$ & $1.12 \times 10^{-15}$ & $3.29 \times 10^{-11}$ \\
\hline 10 GeV & $6.38 \times 10^{-27}$ & $1.31 \times 10^{-24}$ & $2.55 \times 10^{-20}$ & $2.33 \times 10^{-17}$ & $1.43 \times 10^{-11}$ & $5.91 \times 10^{-10}$ \\
\hline 100 GeV & $4.39 \times 10^{-24}$ & $1.27 \times 10^{-23}$ & $2.27 \times 10^{-18}$ & $2.35 \times 10^{-16}$ & $1.22 \times 10^{-12}$ & $6.29 \times 10^{-9}$ \\
\hline 1 TeV & $2.66 \times 10^{-23}$ & $1.22 \times 10^{-22}$ & $6.63 \times 10^{-19}$ & $2.66 \times 10^{-15}$ & $8.32 \times 10^{-11}$ & $8.30 \times 10^{-8}$ \\
\hline \multicolumn{7}{|c|}{$\mathrm{P} 18+\mathrm{BAO}$} \\
\hline 1 MeV & $2.18 \times 10^{-28}$ & $5.32 \times 10^{-26}$ & $9.20 \times 10^{-25}$ & $4.56 \times 10^{-21}$ & $2.33 \times 10^{-19}$ & $1.17 \times 10^{-16}$ \\
\hline 10 MeV & $6.74 \times 10^{-30}$ & $8.29 \times 10^{-26}$ & $4.37 \times 10^{-21}$ & $4.64 \times 10^{-20}$ & $3.12 \times 10^{-17}$ & $1.71 \times 10^{-14}$ \\
\hline 100 MeV & $5.07 \times 10^{-30}$ & $1.35 \times 10^{-25}$ & $3.33 \times 10^{-21}$ & $7.11 \times 10^{-19}$ & $6.58 \times 10^{-16}$ & $1.56 \times 10^{-12}$ \\
\hline 1 GeV & $8.74 \times 10^{-30}$ & $3.13 \times 10^{-25}$ & $5.33 \times 10^{-19}$ & $7.88 \times 10^{-18}$ & $2.51 \times 10^{-12}$ & $1.05 \times 10^{-10}$ \\
\hline 10 GeV & $3.99 \times 10^{-29}$ & $2.36 \times 10^{-24}$ & $1.16 \times 10^{-18}$ & $7.95 \times 10^{-17}$ & $2.63 \times 10^{-13}$ & $1.42 \times 10^{-9}$ \\
\hline 100 GeV & $6.98 \times 10^{-27}$ & $1.47 \times 10^{-23}$ & $6.59 \times 10^{-20}$ & $1.08 \times 10^{-15}$ & $1.15 \times 10^{-11}$ & $1.72 \times 10^{-8}$ \\
\hline 1 TeV & $2.74 \times 10^{-25}$ & $1.81 \times 10^{-22}$ & $1.13 \times 10^{-18}$ & $9.62 \times 10^{-15}$ & $4.83 \times 10^{-11}$ & $1.23 \times 10^{-7}$ \\
\hline
\end{tabular}
\caption{The best-fit values and $95\%$ upper limits obtained by PL for the DM-proton scattering cross section for different combination of
data sets.}
\label{tab:planck_table}
\end{table}

\begin{table}
\begin{tabular}{|c|c|c|c|c|c|}
\hline
\multicolumn{2}{|c|}{\multirow{2}{*}{Parameters}} & 
\multicolumn{2}{c|}{P18+BAO} & \multicolumn{2}{c|}{P18+BAO+DES} \\ \cline{3-6}
\multicolumn{1}{|c}{} & \multicolumn{1}{c|}{} & \multicolumn{1}{c|}{Best-fit} & \multicolumn{1}{c|}{Mean$\pm 1\sigma$} & \multicolumn{1}{c|}{Best-fit} &  \multicolumn{1}{c|}{Mean$\pm 1\sigma$} \\ \hline
\multicolumn{2}{|c|}{$100\Omega_b h^2$} & 2.24 & $2.24_{-0.0136}^{+0.0135}$ & 2.25 & $2.24_{-0.0132}^{+0.0132}$ \\ 
\multicolumn{2}{|c|}{$\Omega_{\rm cdm} h^2$} & 0.119 & $0.120_{-0.000943}^{+0.000936}$ & 0.120 & $0.120_{-0.000913}^{+0.000918}$ \\ 
\multicolumn{2}{|c|}{$100 \theta_s$} & 1.04 & $1.04_{-0.000293}^{+0.000289}$ & 1.04 & $1.04_{-0.000286}^{+0.000286}$ \\ 
\multicolumn{2}{|c|}{$\ln \left(10^{10} A_s\right)$} & 3.06 & $3.05_{-0.0141}^{+0.0141}$ & 3.04 & $3.05_{-0.0141}^{+0.0141}$ \\
\multicolumn{2}{|c|}{$n_s$} & 0.965 & $0.967_{-0.00368}^{+0.000374}$ & 0.971 & $0.966_{-0.00372}^{+0.00371}$ \\
\multicolumn{2}{|c|}{$\tau_{\text {reio}}$} & 0.0598 & $0.0568_{-0.00710}^{+0.000710}$ & 0.0504 & $0.0565_{-0.00715}^{+0.00714}$ \\ \hline
\multicolumn{2}{|c|}{$\sigma_0$/1e-26} & \multicolumn{1}{c|}{0.0218} & 1.85$_{-1.43}^{+1.49}$ & \multicolumn{1}{c|}{0.0919} & 0.993$_{-0.711}^{+0.621}$\\ \hline
\multicolumn{2}{|c|}{$\Omega_m$} & 0.309 & $0.310_{-0.000570}^{+0.000568}$ & 0.309 & $0.312_{-0.00558}^{+0.00556}$ \\
\multicolumn{2}{|c|}{$H_0$} & 67.7 & $67.6_{-0.424}^{+0.423}$ & 67.8 & $67.5_{-0.412}^{+0.412}$ \\
\multicolumn{2}{|c|}{$S_8$} & 0.827 & $0.823_{-0.0111}^{+0.0115}$ & 0.820 & $0.827_{-0.0109}^{0.0109}$ \\ \hline
\multicolumn{1}{|c|}{\multirow{2}{*}{$-2\ln\mathcal{L}$}} & \multicolumn{1}{c|}{DES} & \multicolumn{2}{c|}{-} & \multicolumn{2}{c|}{518.4} \\ \cline{2-6} 
& \multicolumn{1}{c|}{Total} & \multicolumn{2}{c|}{2785.16} & \multicolumn{2}{c|}{3303.32}
\\ \hline

\end{tabular}
\caption{The best-fit parameter values 
 for benchmark parameter $n=0$ and $m_\chi=1\mev$.}
\label{tab:bf_value}
\end{table}

Incorporating DES \texttt{3$\times$2pt} data improves constraints compared to early universe observations (P18 + BAO) by a factor ranging from one to five.
Our results support conclusions from previous studies~\cite{Maamari:2020aqz,Zu:2023rmc} that observations of WL on small-scale matter distributions in the late universe provide a more precise exploration of DM physics.
Finally, the projected $95\%$ upper limits (the green lines in Fig.\ref{fig:upperlimits}) are derived from the CSST mock cosmic shear data with PL method. 
We generated the CSST mock cosmic shear data and the corresponding covariance matrix, 
employing CSST sensitivity settings based on P18 $\Lambda$CDM parameters: 
\{{$\Omega_{\rm b} h^2=0.02238$, $\Omega_{\rm cdm} h^2=0.1201$, $\theta_{\rm s}=0.01041$, $\log(10^{10} A_{\rm s})=3.045$, $n_{\rm s}=0.9659$, $\tau_{\rm reio}=0.0543$}\}. 
The covariance matrix was computed using the \texttt{CosmoCov} code~\cite{Fang:2020vhc,Krause:2016jvl}, 
incorporating a conservative galaxy shape noise of $\sigma_e=0.3$~\cite{Miao:2022hyp} and the red-shift distributions of mock CSST galaxies from Ref.~\cite{Lin:2022aro}. 
Again, we applied the same small angular scale mask as DES does in Appendix~\ref{WL}. 
Remarkably, the future CSST cosmic shear sensitivity may substantially lower down the upper limit by around one to two orders compared to the DES \texttt{3$\times$2pt} likelihood. 
Note that the primary theoretical uncertainties in our study arise from baryon feedback and the interpolation range of our PCA machine about percent level, see Appendix~\ref{error} for details. 
We expect to address and refine these aspects in future work to enhance the overall robustness of our results.

\begin{figure}[ht!]
    \centering
    \includegraphics[scale=0.35]{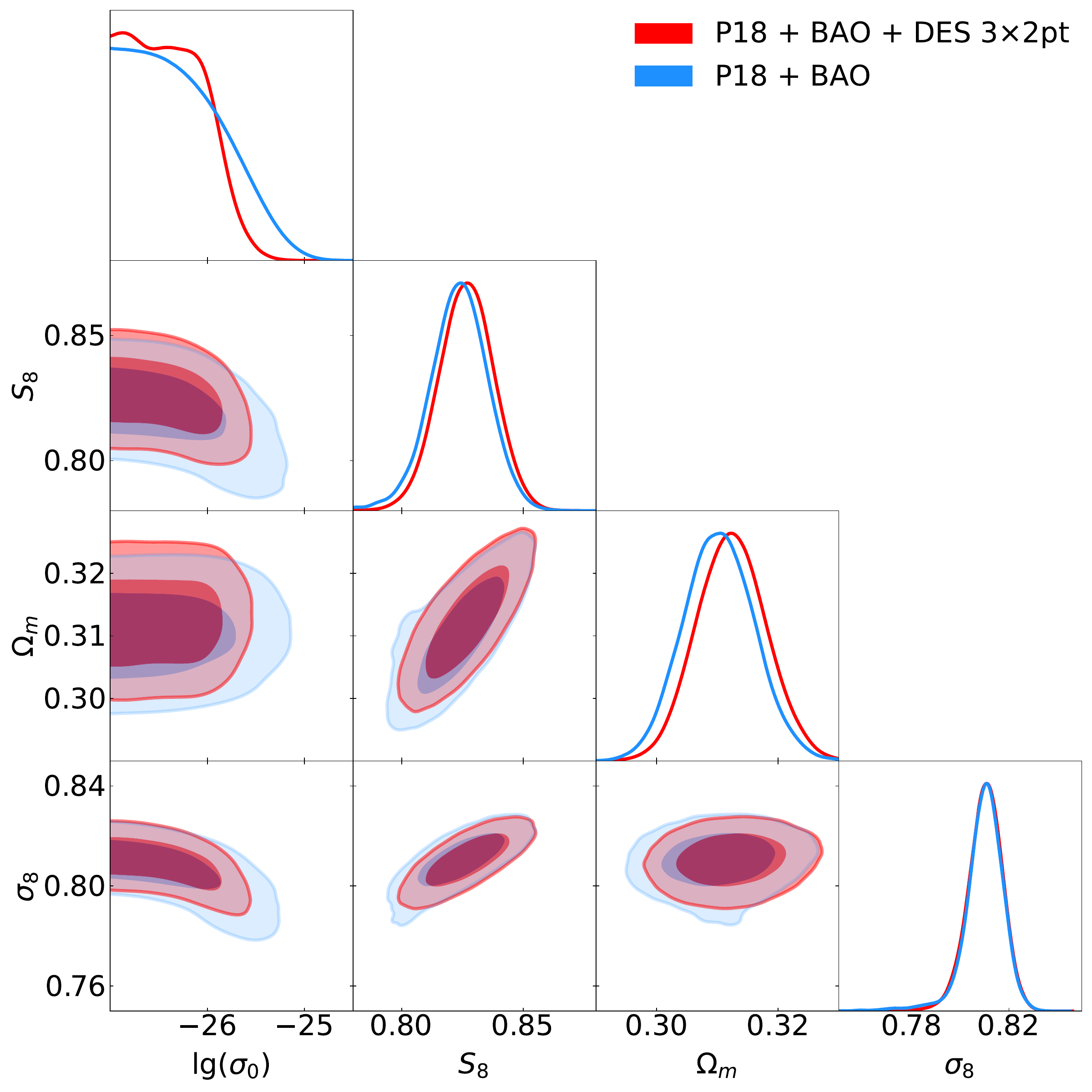}
    \caption{The two dimensional marginalized posterior distribution of DM-proton scattering cross section $\lg(\sigma_0)$, the matter fluctuation $S_8$, the total matter density $\Omega_m$ and $\sigma_8$ for different combination of data sets.
    }
    \label{fig:n0_1mev_triangle}
\end{figure}

In Fig.~\ref{fig:n0_1mev_triangle}, assuming $m_\chi = 1\mev$, we present the two-dimensional marginalized posterior distribution of four parameters:
$\sigma_0$, total matter density $\Omega_m$, matter fluctuations parameterized by $S_8$, and $\sigma_8$. 
The plots indicate that the ``large" interactions between DM particles and protons can suppress mass fluctuations, resulting in a decreased $S_8$, potentially alleviating the tension related to $S_8$. 
However, the DES likelihoods, particularly through galaxy-galaxy lensing $\gamma_t$ and galaxy clustering $w$, impose more stringent upper limits on the cross sections, thereby weakening the suppression effect on small-scale structure.

\section{Summary and conclusion}
\label{conclusion}

In order to study the role of DM-proton scattering in cosmology, we focused on the matter distribution at scales $k>0.1\,h/\rm{Mpc}$, where non-linear effects become important. 
Accounting for the impact of DM-proton scattering in the non-linear evolution of structure formation, 
we computed 205 linear matter power spectra as initial conditions at $z=127$ for subsequently conducted cosmological $N$-body simulations. 
To incorporate these effects into the MCMC global scan, 
we utilized the PCA method to establish a mapping between any linear ratio spectrum (the spectrum of the DM-proton scattering scenario to that of $\Lambda$CDM) and the corresponding ratio derived from $N$-body simulations.

Employing this approach, we utilized data from DES Y3 WL \texttt{3$\times$2pt}, together with likelihoods derived from Planck 2018 CMB and BOSS DR12 BAO, to constrain the velocity-dependent elastic scattering cross-sections between DM particles and protons. 
We find null signals of DM-proton scattering, thus we set an upper limit on the cross section.
Our comprehensive studies of the parameter space in the context of DM-proton scattering cosmology allowed us to estimate the $95\%$ upper limits of cross-sections $\sigma_{n=0,2,4}$ (refer to Fig.~\ref{fig:upperlimits}). 
It is worthy to mention that the constraints derived from the DES \texttt{3$\times$2pt} data showed a substantial improvement, achieving a factor of $\mathcal{O}(5)$ over previous results.
Our results, introduced as the first Frequentist upper limits, are compared with those obtained using a Bayesian approach for a more comprehensive understanding.

In the near future, the precision of weak lensing measurements in galaxy surveys is set for further investigation into DM-proton scattering.
To explore the sensitivities of future WL data, we generated cosmic shear mock data and a covariance matrix for the CSST. 
The resulting sensitivities from CSST exhibit a remarkable improvement of one to two orders of magnitudes compared to the current DES \texttt{3$\times$2pt} data, as illustrated in Fig.~\ref{fig:upperlimits}.

Besides two-point correlation functions, the analysis of higher-order statistics that contain cosmological information beyond that captured by two-point correlation functions, may further improve the limits of the cross-section~\cite{Takada:2003sv,Halder:2021itp,Halder:2023kfy,Boyle:2020bqn,Barthelemy:2023mer,DES:2019ujq,DES:2021lsy}. 
This can be an interesting avenue to explore in future works.
Based on our theoretical framework as well as the simulations for DM-proton scattering modelling, 
we only perform the cosmological analyses with two point statistics in this work, 
but would like to return to high-order statistics in the future.

\section{Acknowledgements}
We are grateful to Rui An, Zi-qing Xia, Wentao Luo, Yin Li, Yuchao Gu, Lei Lei for useful discussions. 
This work is supported by the National Key Research and Development Program of China (No. 2022YFF0503304), CSST science funding, 
and the Project for Young Scientists in Basic Research of the Chinese Academy of Sciences (No. YSBR-092). 

\appendix

\section{Settings of $N$-body simulations}
\label{nbody_set}
We briefly introduce the settings of our $N$-body simulations.
In this work, we conducted our $N$-body simulations using the TreePM code \texttt{GIZMO}~\cite{Hopkins:2014qka}, 
where the $N$-body part originates from the source code of \texttt{Gadget2}~\cite{Springel:2005mi}.
Our simulations start at $z = 127$,  with the initial conditions generated by the code \texttt{2LPTic}~\cite{Crocce:2006ve}. 
The input power spectrum and corresponding $\sigma_8$ were obtained from \texttt{CLASS}.
For all simulation points, we utilized a set of fiducial cosmological parameters \{$\Omega_m=0.3118$, $\Omega_\Lambda=0.6881$, $h=0.6760$, $n_s=0.9659$\}.

\section{Error estimation of our approach}
\label{error}
\begin{figure}[h!]
    \centering
    \includegraphics[scale=0.4]{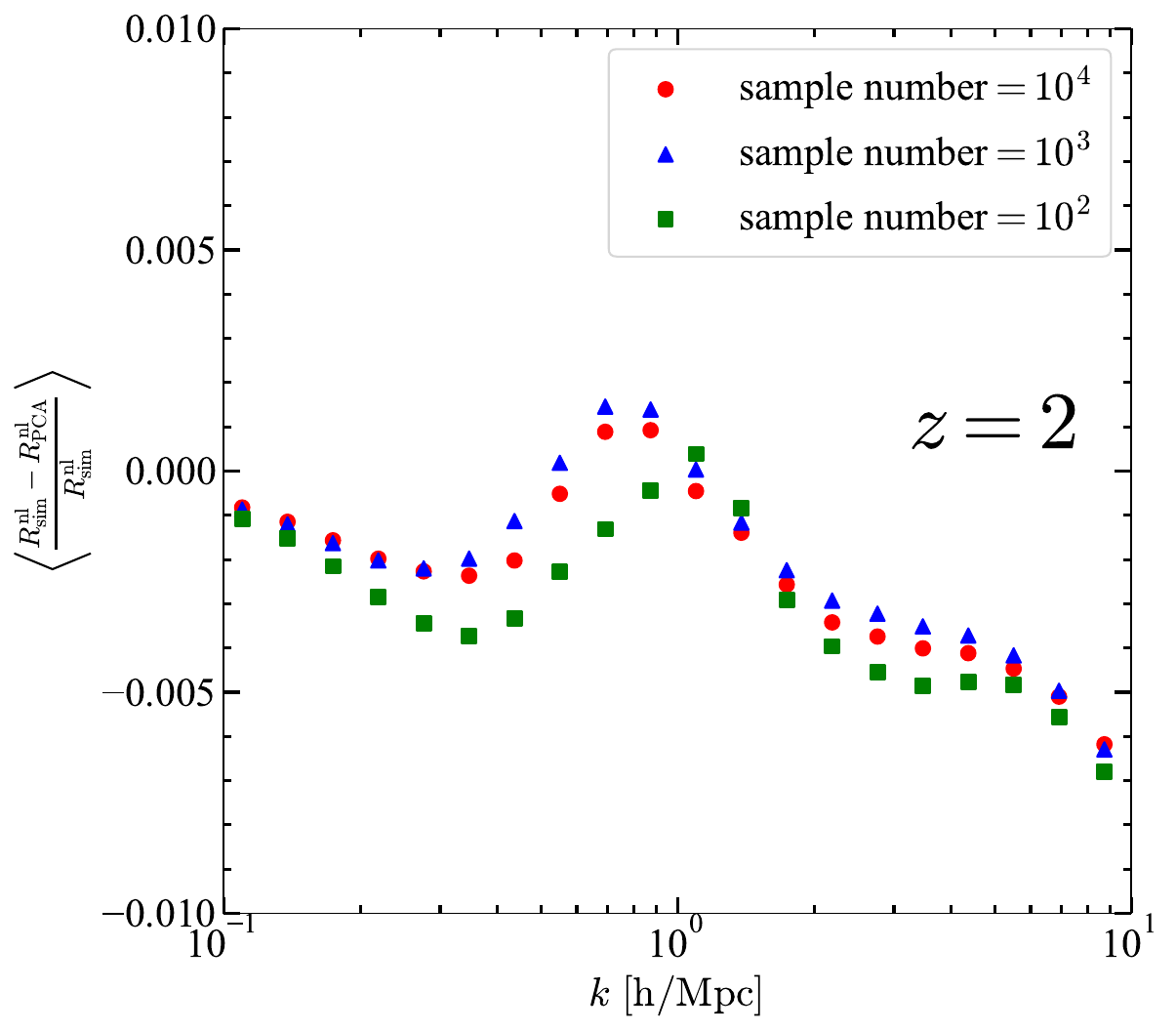}
    \caption{The average non-linear ratio deviation between simulation results and our approach output at each wave number ($k$) at $z=2$. The green cube, blue triangle and red dot marker correspond to sample number $N=10^2,10^3,10^4$, respectively.
    }
    \label{fig:simu_dmp_err}
\end{figure}
We test the robustness of our approach.
Firstly, we randomly take 5 test points from the total 205 simulation points and using the left 200 points to build the interpolation map.
Then we compute the average non-linear ratio deviation between the output of our approach (PCA) and simulation (sim) at each wave number ($k$) bins for the 5 test points, as once sampling.
Finally, we compute the total average non-linear ratio deviation by summing over all the sample numbers,
\begin{align}
    \left<\frac{R^{\rm nl}_{\rm sim}-R^{\rm nl}_{\rm PCA}}{R^{\rm nl}_{\rm sim}}\right> = \frac{1}{N}\sum\limits_{N} \frac{1}{5}\sum\limits_{i=1}^{i=5}\frac{R^{\rm nl}_{{\rm sim}, i}(k,z)-R^{\rm nl}_{{\rm PCA}, i}(k,z)}{R^{\rm nl}_{{\rm sim}, i}(k,z)},
\end{align}
where $N$ is the sample number.
We find the largest deviation appeared at $z=2$ is shown in Fig.~\ref{fig:simu_dmp_err}, which prove our approach could provide valid non-linear correction with an error less than $0.5\%$ in DM-proton scattering scenario.

\section{The supplementary results for P18+BAO+DES $\xi_\pm$ likelihood}
\label{CS}

\begin{figure}[h!]
    \centering
    \includegraphics[scale=0.35]{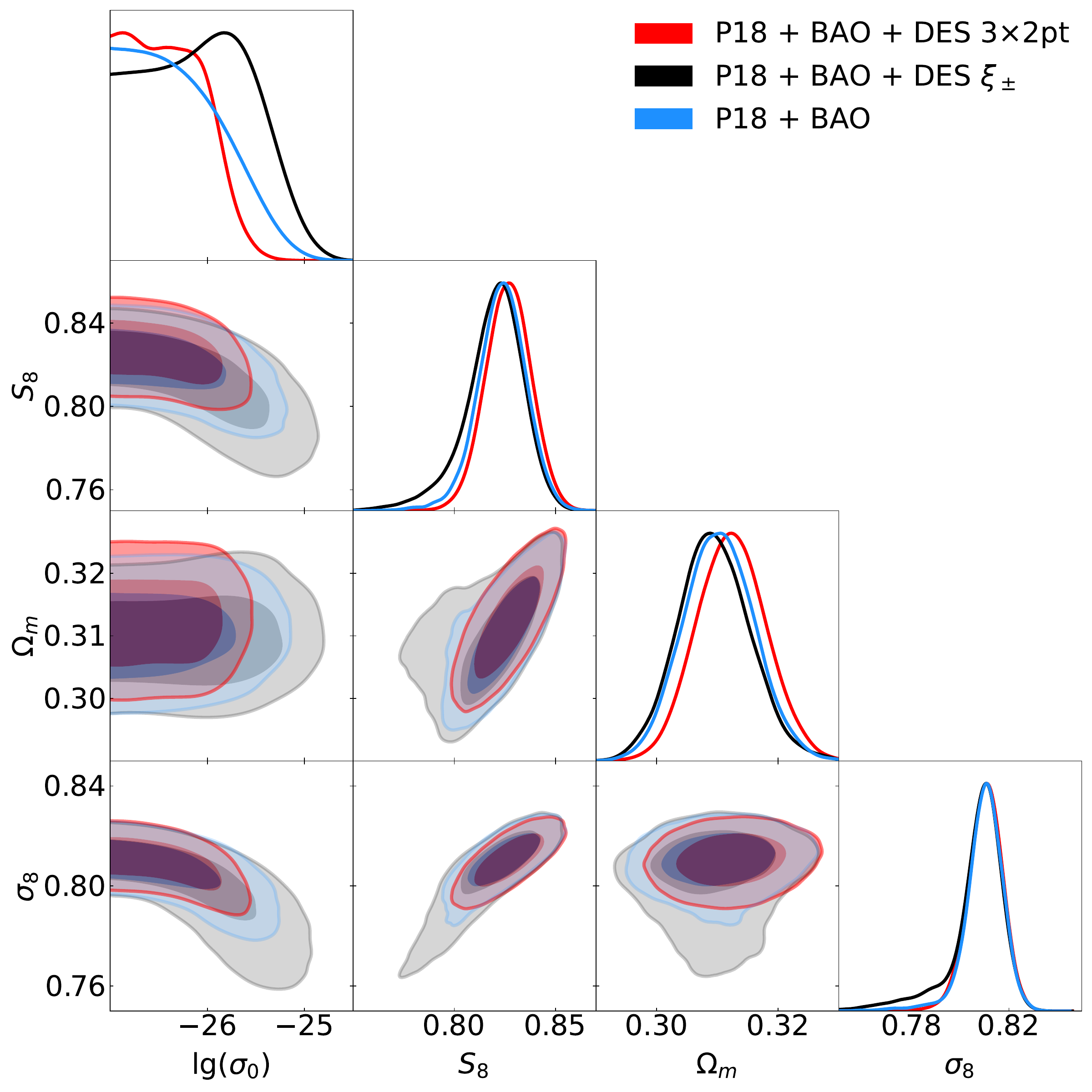}
    \caption{Similar plot as Fig.~\ref{fig:n0_1mev_triangle}, 
    but including the likelihood from P18 + BAO + DES $\xi_\pm$ data set.
    }
    \label{fig:shearonly_results}
\end{figure}

The $\gamma_t$ and $w$ are the biased tracers of matter field, 
which may be contaminated by the systematical uncertainties from the galaxy bias.    
To a complementary point of view, we also perform a scan using P18 + BAO + DES cosmic shear likelihood, by taking $n=0$ and $m_\chi=1\mev$. 
In Fig.~\ref{fig:shearonly_results}, we present the two dimensional marginalized posterior distribution of $\sigma_0$, $S_8$, $\Omega_m$, and $\sigma_8$, by considering three sets of likelihoods: 
P18+BAO+DES 3$\times$2pt (red), 
P18+BAO+DES $\xi_{\pm}$ (black), 
and P18+BAO (blue).

When comparing the two likelihoods (P18+BAO and P18+BAO+DES $\xi_{\pm}$), 
we find that cosmic shear data (DES $\xi_{\pm}$) slightly favors a larger cross-section region, indicating a suppression in mass fluctuations (also see Fig.~\ref{fig:shear_pm_tot}). 
However, incorporating $\gamma_t$ and $w$ into the likelihood, the cross-section tends to favor smaller values.

\section{The DES modeling and data}
\label{WL}
In this section, we outline the WL observables and model predictions according to reference~\cite{DES:2021wwk,DES:2021bvc,DES:2021vln,Krause:2016jvl,Krause:2015jqa}. 
Additionally, we illustrate the dependency of WL modeling on the non-linear matter power spectrum derived in Sec.~\ref{methodology}.

We start our analysis with the DES~Y3 WL data employed in this study. 
This data-set encompasses measurements of galaxy shapes and distributions across $4143\,{\rm deg}^2$ of the entire sky.
The observed galaxies were sorted into two catalogs: the \texttt{METACALIBRATION}~\cite{DES:2020ekd} source catalog and the \texttt{MagLim}~\cite{DES:2020ajx} lens catalog.
The source catalog contains $10^8$ galaxies and is categorized into four tomographic red-shift bins with bin edges at $z = [0.0, 0.36, 0.63, 0.87, 2.0]$, 
while the lens catalog encompasses $10^7$ galaxies, separated into six tomographic red-shift bins with bin edges at $z = [0.20, 0.40, 0.55, 0.70, 0.85, 0.95, 1.05]$. 
Each redshift bin $i$ or $j$ includes a red-shift density distribution of galaxies, denoted as $n_\kappa^i(z)$ for \texttt{METACALIBRATION} and $n_g^j(z)$ for \texttt{MagLim}. 
The measured shape distortions and angular positions of galaxies provide three sets of two-point correlations, referred to as `\texttt{3$\times$2pt}'.
These correlations include the angular separations of foreground lens galaxy pairs, measured through galaxy clustering $w(\theta)$ (position-position).
The shape distortions of source galaxies can also be related to the distribution of foreground lens galaxies, measured via galaxy-galaxy lensing $\gamma_t(\theta)$ (position-shape). 
However, since galaxies act as a biased tracer of the matter field, 
cosmic shear $\xi_{\pm}(\theta)$ (shape-shape), 
as a cosmological WL method involves correlations between the shape distortions of source galaxy pairs, 
directly trace the foreground matter field and is extremely sensitive to the LSS, as well as the matter power spectrum in the late universe.
The DES collaboration offers a comprehensive WL \texttt{3$\times$2pt} data~\footnote{\label{desy3data}\url{https://des.ncsa.illinois.edu/releases/y3a2/Y3key-products}}, encompassing both auto and cross-correlations among different red-shift bins.
Each observable is computed across $20~\theta$-bins logarithmically spanning from $2.5$ to $250$ arcmin, along with its respective covariance matrix.
However, the linear galaxy bias assumption breaks down on small angular scales, and complexities in baryonic physics introduce uncertainties.
We adopt a mask method for the small angular scale data, consistent with~\cite{DES:2021wwk}.
Additionally, due to the poor model fit caused by high $z$ bins data in \texttt{MagLim}, we implement a high $z$ cut, resulting in the exclusion of bins $5$ and $6$ from the \texttt{MagLim} in our analysis~\cite{DES:2021wwk}.
It's important to note that the complete \texttt{3$\times$2pt} data vector consists of 1000 elements, only 462 elements remain after masking.

Now, we elucidate the model predictions for \texttt{3$\times$2pt} signals. 
Utilizing the Limber approximation~\cite{Limber:1953,LoVerde:2008re}, we can compute the angular power spectrum for the \texttt{3$\times$2pt} observable between red-shift bins $i$ and $j$:
\begin{equation}
    C^{ij}_{\rm \kappa \kappa, \kappa g,gg}(\ell) = \int {\rm d} \chi \frac{q^i_{\rm g/\kappa} q^j_{\rm g/\kappa}}{\chi^2} \times P_{\rm \chi p}^{\rm nl}\left(k=\frac{\ell +0.5}{\chi}, z(\chi)\right),
\end{equation}
where $\kappa$ and $\mathrm{g}$ denote the source field and lens field, respectively. The symbol $\chi$ represents the comoving distance, while $\ell$ stands for the angular wave number. The 3D non-linear matter power spectrum, incorporating the characteristics of DM-proton scattering denoted as $P_{\rm \chi p}^{\rm nl}$, was derived in Eq.~\ref{halofit_times_ratio}.
The radial weight function for clustering in red-shift bin $i$ is
\begin{equation}
    q_g^i(\chi) = b^in_g^i(z(\chi))\frac{{\rm d}z}{{\rm d}\chi}
\end{equation}
and the lensing efficiency kernel in red-shift bin $j$ is
\begin{equation}
q_\kappa^j(\chi)=\frac{3 H_0^2 \Omega_m}{2 \mathrm{c}^2} \frac{\chi}{a(\chi)} \int_\chi^{\chi_{\mathrm{z_{max}}}} \mathrm{d} \chi^{\prime} n_\kappa^j\left(z\left(\chi^{\prime}\right)\right) \frac{\mathrm{d} z}{\mathrm{d} \chi^{\prime}} \frac{\chi^{\prime}-\chi}{\chi^{\prime}},
\end{equation}
where $b^i$ represents the linear galaxy bias within redshift bin $i$. 
The normalized distributions of lens galaxies in red-shift bin $i$ and source galaxies in bin $j$ correspond to $n_g^i(z)$ and $n_\kappa^j(z)$ respectively, which can be obtained from DES~Y3 products~\footref{desy3data}. 
The parameters $H_0$, $\Omega_m$, $c$, and $a(\chi)$ signify the Hubble constant, total matter density, speed of light, and scale factor. 
The angular correlation function for galaxy clustering can be computed from
\begin{equation}
w^i(\theta)= \int \frac{\mathrm{d} \ell \ell}{2 \pi} J_0(\ell \cdot \theta) C_{\mathrm {gg}}^{ij}(\ell),
\end{equation}
the galaxy-galaxy lensing correlation function for lens galaxies in bin $i$ and source galaxies in bin $j$ can be writen as 
\begin{equation}
\gamma_{\mathrm{t}}^{i j}(\theta)= \left(1+m^j\right) \int \frac{\mathrm{d} \ell \ell}{2 \pi} J_2(\ell \cdot \theta) C_{\mathrm {g\kappa}}^{ij}(\ell),
\end{equation}
and the cosmic shear correlation functions are
\begin{equation}
\xi_{\pm}^{i j}(\theta)= \left(1+m^i\right)\left(1+m^j\right) \int \frac{\mathrm{d} \ell \ell}{2 \pi} J_{0/4}(\ell \cdot \theta) C_{\mathrm {\kappa \kappa}}^{ij}(\ell),
\end{equation}
where $J_0$, $J_2$, and $J_4$ denote the zeroth-order, second-order, and fourth-order Bessel functions, respectively. 
The pre-factor $m^i$ represents the multiplicative shear bias in bin $i$.
Additionally, several other critical systematic uncertainties must be taken into account. 
For instance, the red-shift distributions of galaxies in bin $i$ need calibration due to photo-$z$ uncertainties $\Delta z^i$, adjusted as $n^i(z-\Delta z^i)$.
The intrinsic alignment (IA) of source galaxies significantly impacts $\gamma_t^{ij}(\theta)$ and $\xi_\pm^{ij}(\theta)$. 
We address this issue using a non-linear alignment (NLA) model~\cite{Krause:2015jqa,HSC:2018mrq}.

\begin{figure}[h!]
    \centering
    \includegraphics[scale=0.4]{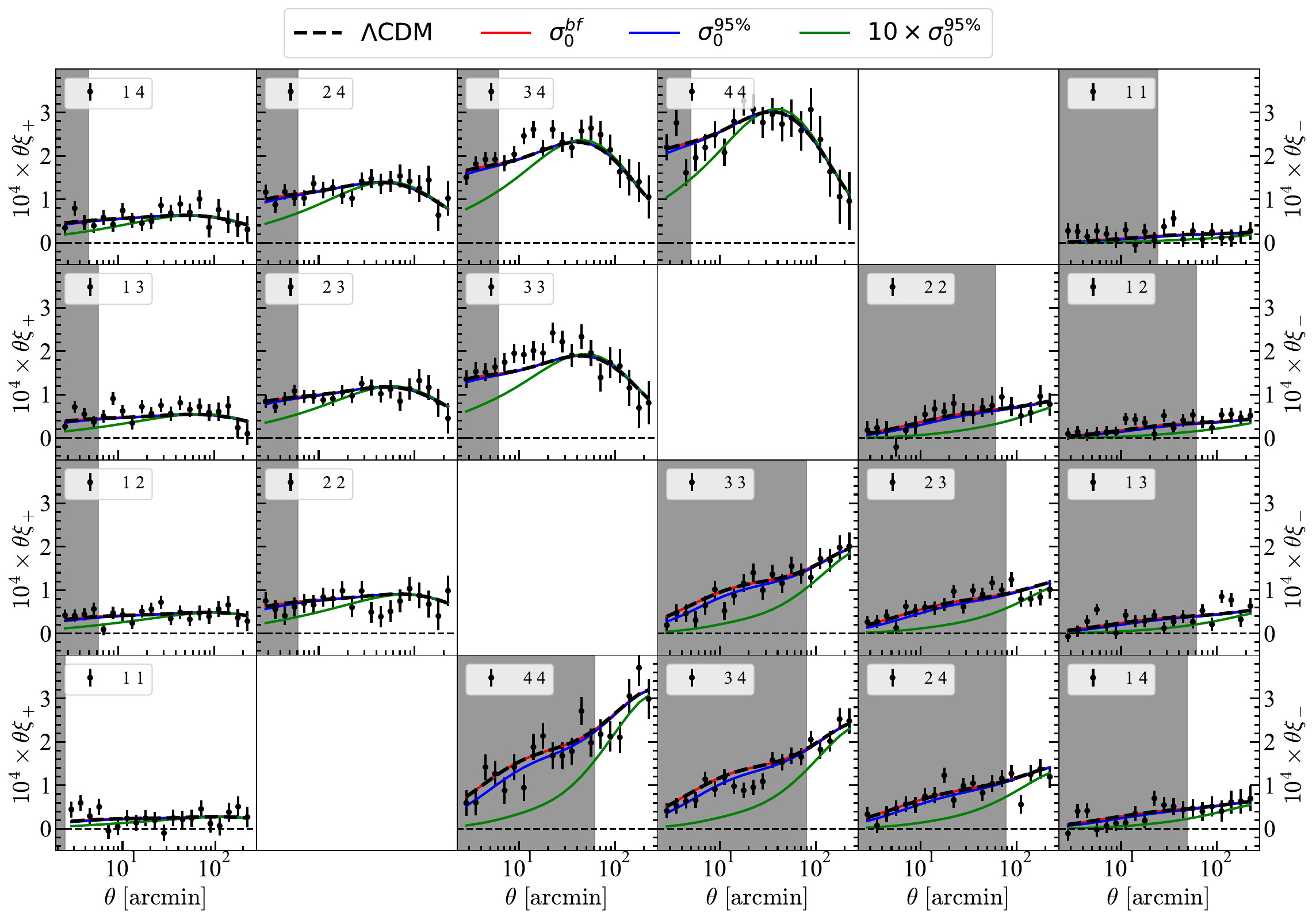}
    \caption{The cosmic shear correlation function across all combination of source red-shift bins. The $\xi_+$ are on the left while $\xi_-$ on the right. The gray shaded region is masked due to the small scale uncertainties.
    }
    \label{fig:shear_pm_tot}
\end{figure}

\begin{figure}[h!]
    \centering
    \includegraphics[scale=0.4]{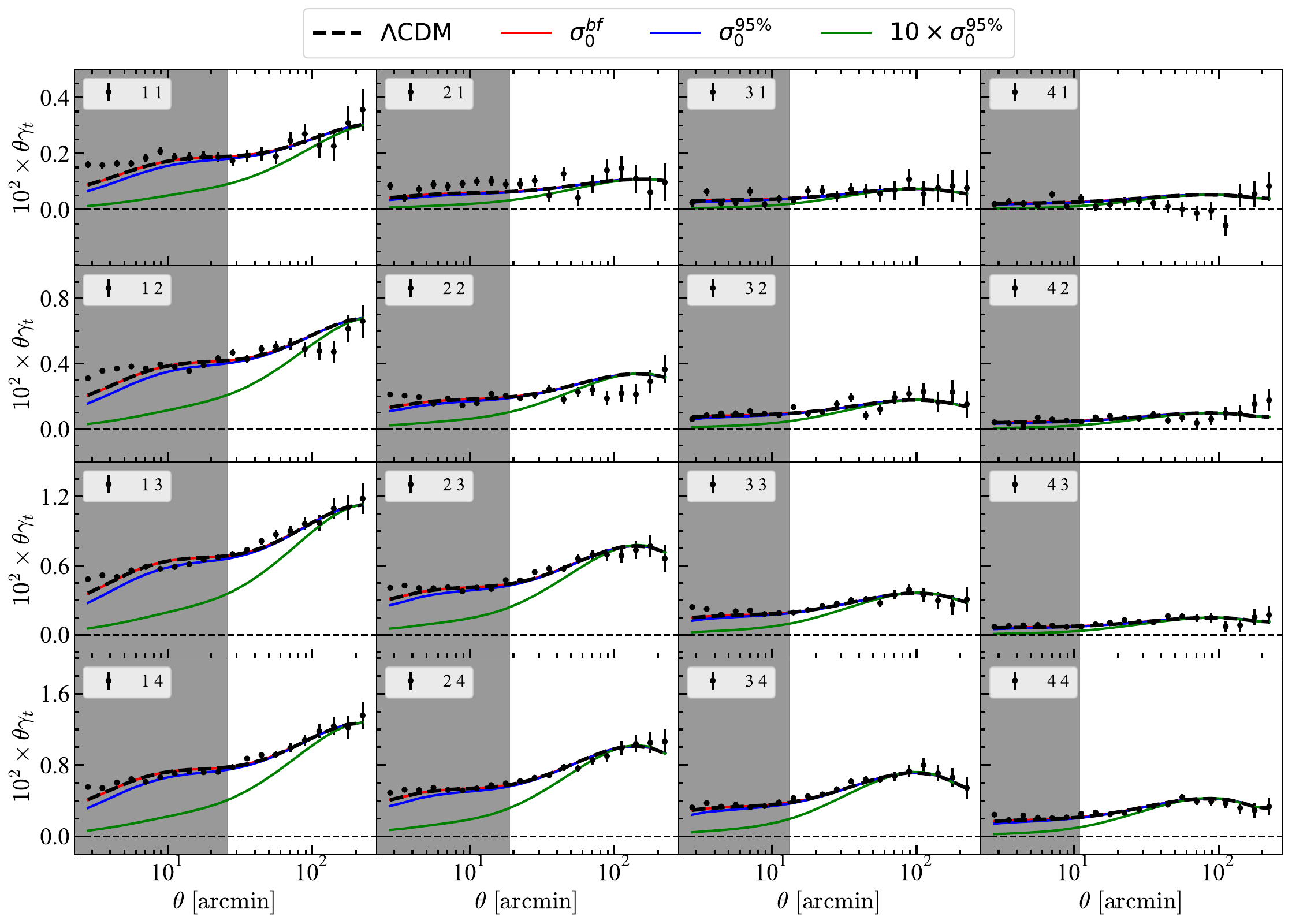}
    \caption{Similar plot as Fig.~\ref{fig:shear_pm_tot} but for galaxy-galaxy lensing $\gamma_t$ across all of the source and lens red-shift bins. 
    }
    \label{fig:GG_tot}
\end{figure}

\begin{figure}[h!]
    \centering
    \includegraphics[scale=0.4]{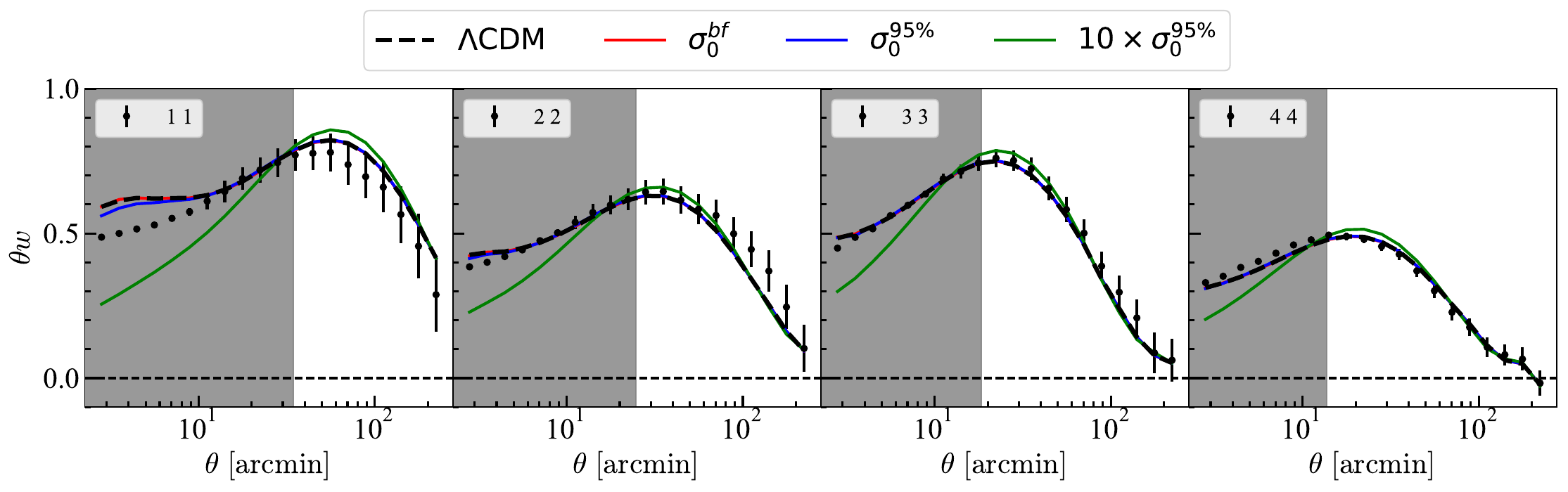}
    \caption{Similar plot as Fig.~\ref{fig:shear_pm_tot} but for galaxy clustering $w$ across all of the lens red-shift bins. 
    }
    \label{fig:clustering_tot}
\end{figure}

We present the full set of 3$\times$2pt plots here with both data and model predictions similar to Fig.~\ref{fig:3x2pt_fitting}, including all auto and cross correlations across different red-shift bins.
The cosmic shear $\xi_\pm$, galaxy-galaxy lensing $\gamma_t$, galaxy clustering $w$ correspond to the Fig.~\ref{fig:shear_pm_tot}, Fig.~\ref{fig:GG_tot} and Fig.~\ref{fig:clustering_tot}, respectively.

\bibliographystyle{JHEP}
\bibliography{bibtex.bib}
\end{document}